\documentclass[a4paper,11pt]{article}
\usepackage[top=1.8cm,bottom=1.8cm,left=1.8cm,right=1.8cm,asymmetric]{geometry}
\usepackage{url}
\usepackage{paralist}
\usepackage{enumitem}
\usepackage[noadjust]{cite}
\usepackage[hyphenbreaks]{breakurl}
\usepackage{authblk}
\usepackage{tabularx}
\usepackage{booktabs} 
\usepackage[colorlinks=true,hyperfootnotes=true]{hyperref}

\title{A UK Case Study on\\Cybersecurity Education and Accreditation}

\author[1]{Tom Crick}
\author[2]{James H. Davenport}
\author[3]{Alastair Irons}
\author[4]{Tom Prickett}
\affil[1]{Swansea University, Swansea, UK}
\affil[2]{University of Bath, Bath, UK}
\affil[3]{Sunderland University, Sunderland, UK}
\affil[4]{Northumbria University, Newcastle upon Tyne, UK}
\affil[1]{\url{thomas.crick@swansea.ac.uk}}
\affil[2]{\url{j.h.davenport@bath.ac.uk}}
\affil[3]{\url{alastair.irons@sunderland.ac.uk}}
\affil[4]{\url{tom.prickett@northumbria.ac.uk}}

\date{17 July 2019}

\begin{document}
\maketitle

\begin{abstract}
This paper presents a national case study-based analysis of the numerous dimensions to cybersecurity education and how they are prioritised, implemented and accredited; from understanding the interaction of hardware and software, moving from theory to practice (and vice versa), to human factors, policy and politics (as well as various other important facets). A multitude of model curricula and recommendations have been presented and discussed in international fora in recent years, with varying levels of impact on education, policy and practice. This paper address three key questions: {\emph{i)}} what is taught and what should be taught for cybersecurity to general computer science students; {\emph{ii)}} should cybersecurity be taught stand-alone or in an integrated manner to general computer science students; and {\emph{iii)}} can accreditation by national professional, statutory and regulatory bodies enhance the provision of cybersecurity within a body's jurisdiction?

Evaluating how cybersecurity is taught in all aspects of computer science is clearly a task of considerable size, one that is beyond the scope of this paper. Instead a case study-based research approach -- primarily focusing on the UK -- has been adopted to evaluate the evidence of the teaching of cybersecurity within general computer science to university-level students. Thus, in the context of widespread international computer science/engineering curriculum reform, what does this need to embed cybersecurity knowledge and skills mean more generally for institutions and educators, and how can we teach this subject more effectively? Through this UK case study, and by contrasting with related initiatives in the US, we demonstrate the positive effect that national accreditation requirements can have, and offer some recommendations both for future research and curriculum developments.
\end{abstract}

\noindent {\footnotesize{{\textbf{Keywords}}: Cybersecurity, curricula, accreditation, computer science education, public policy, UK}}

\section{Introduction}

Cybersecurity has increasingly been a headline feature in the news over recent years, generally prompted by spectacular breaches, including major credit reference agencies~\cite{equifax2017}, telecoms companies~\cite{talktalk2016}, national airlines~\cite{BritishAirways2018a}, online dating websites~\cite{ashleymadison2015}, and even between sovereign governments~\cite{ncsc2018}. These major breaches have had significant impact on both individual citizens and society in general, requiring attention from organisations of all sizes:

\begin{quote}
``{\emph{...[need to] change the culture in your organisation around cyber security; to try to do for cyber what has been done so successfully for health and safety, for example, over the last ten years --- to get everybody to take it seriously; to take the risk management process seriously and drive that down through the organisation.}}'' Robert Hannigan~\cite{Hannigan2019a}, former Director of GCHQ
\end{quote}

These global cybersecurity crises have compelled academic institutions to address the demand for educated cybersecurity professionals~\cite{McGettrick2013}. As no shared framework for ``cybersecurity'' as an academic discipline exists, growth has been unfocused and largely driven by training materials, which makes it harder to establish a common body of knowledge (for example, in the UK, the CyBOK project~\cite{Bristol2019a} is still a work in progress). An international perspective is harder still, as different nations use different criteria to define local needs~\cite{schneider2013}. As a result, new programmes entering this space are free to conceptualise, design, package and market their initiatives, as there is no globally accepted reference model for cybersecurity to allow employers or students to understand the extent or ambition of a given cybersecurity program~\cite{conklin-et-al:2014,Parrishetal2018a}.

With this significant economic and societal focus on cybersecurity, there are calls for formal education -- school-level as well as tertiary -- to respond to this situation, at the individual level and via recommended curricula~\cite{mcgettrick-et-al:sigcse2014,ACM2017b} and professional accreditation requirements~\cite{BCS2018a,NCSC2017}. This is further reinforced by a wider focus on digital skills and computer science education reform, especially across the nations of the UK~\cite{brown-et-al:toce2014,murphy-et-al:programming2017,tryfonas+crick:petra2018}. 

An ACM working group~\cite{Parrishetal2018a}, established in 2018 as part of the {\emph{Innovation and Technology in Computer Science Education}} (ITiCSE) conference series, has been capturing global perspectives on cybersecurity education, but as of July 2019 has yet to report its full findings. An aim of the ITiCSE working group is to develop a taxonomy for approaches to cybersecurity education, resulting in improved standards and goals for many different types of cybersecurity programmes; a further aim is to ``{\emph{catalog existing [\dots] knowledge materials}}'', but there is no mention of any quality control over these. Nevertheless, it is one thing to write national curricula, specifications and requirements, and another thing to deliver appropriate and relevant education and skills; this paper asks how well this is done in practice.

\section{Cybersecurity For All, Or Just For Specialists?}

In one sense, this title is a false dichotomy: there is a serious need for cybersecurity specialists (estimates vary, but are always large), but also all in IT need to know \emph{some} cybersecurity -- thus, there is a case for depth as well as breadth~\cite{manson+pike:2014,davenport-et-al:latice2016}. This is not a new concern~\cite{Parr2014a}, but it is a growing one.

This need to build knowledge, skills and capacity in the area of cybersecurity has also led to the establishment of a number of strategic policy initiatives from a number of national governments, for example the publication in 2016 of the UK's Cyber Security Strategy~\cite{ukcyberstrategy:2016} (along with the setting up of the National Cyber Security Centre, as well as increased scrutiny of the resilience of the UK's critical national infrastructure~\cite{lordscyberreport:2018}; also industry-focused initiatives such as {\emph{Cyber Essentials}}\footnote{{\emph{Cyber Essentials}} is a UK Government-backed, industry-supported scheme to help organisations protect themselves against common online threats: \url{https://www.cyberessentials.ncsc.gov.uk}}), the EU Cybersecurity Act~\cite{eucyber2018} (which reinforces the mandate of the EU Agency for Cybersecurity: ENISA, the European Union Agency for Network and Information and Security), or the National Initiative for Cybersecurity Education (NICE) in the USA~\cite{NICE}.

The teaching of cybersecurity in higher education pre-dates these initiatives and there has been recognition of the need for the inclusion of cybersecurity as part of the discipline of computer science for a number of years~\cite{Hentea2006,schneider2013}. There have been a number of international initiatives to define curricula to support this, for example the 2013 ACM Computer Science Curricula Recommendations~\cite[which added ``{\emph{Information Assurance and Security}}'' for the first time]{ACM2013a}, as well as for specialised cybersecurity degree programmes~\cite{ACMIEEEAISSIGSECIFIP}. There has been debate as to whether cybersecurity is distinct discipline from computer science \cite{McGettrick2013}; the consensus increasingly is that cybersecurity is both a discipline in its own right and that cybersecurity should be taught within computer science and related degrees. Recent evolution of cybersecurity education shows that it has begun to take shape as a true academic perspective, as opposed to simply being a training domain for certain specialised jobs. More recent work presents cybersecurity as a ``meta-discipline''; that is, cybersecurity should be used as an aggregate label for a wide variety of similar disciplines, much in the same way that the terms ``engineering'' and ``computing'' are commonly used~\cite{Parrishetal2018b}. 

Further to the substantial computer science and digital skills curriculum reform across the UK~\cite{crick+sentance:2011,wgictreview:2013,brown-et-al:toce2014,tryfonas+crick:petra2018}, the question is raised whether cybersecurity should be formally taught in schools, as part of compulsory school-level education. While aspects of ``e-safety'' and principles of protecting personal data are increasingly visible in formal curricula~\cite{moller+crick:jce2018}, the majority of UK schoolchildren's exposure to cybersecurity skills is through national extra-curricular competitions, for example Cyber Security Challenge UK\footnote{\url{https://www.cybersecuritychallenge.org.uk}}. The Institute of Coding, a \pounds40m+ initiative by the UK Government to transform the digital skills profile of the country~\cite{Davenportetal2019a} -- but primarily focused on university graduates -- does indeed mention cybersecurity, but merely as a sub-item in one work package.

\section{Methodology}

\subsection{Research Questions}

There are various levels of specialism at which cybersecurity education and skills can be addressed; for example:

\begin{enumerate}[label=(\roman*)]
\item The generalist computer science graduate;
\item The generalist computer science masters graduate;
\item The specialist computer science graduate;
\item The specialist computer science masters graduate;
\item The reskilling/upskilling/professional development of the IT industry and the wider workforce;
\item The general public --- this is important, but there are many initiatives in this area, which are, rightly, largely separated from computing education.
\end{enumerate}

The focus of this paper is on {\emph{(i)}}--{\emph{(ii)}}: the general computer science graduate. We thus focus on three research questions:

\begin{description}
\item[RQ1] {\emph{What cybersecurity is taught and what cybersecurity should be taught to the general computer science students?}}
\item[RQ2] {\emph{Should cybersecurity be taught stand-alone or in an integrated manner to general computer science students?}}
\item[RQ3] {\emph{Can accreditation by professional, statutory and regulatory bodies (PSRBs) enhance the provision of cybersecurity within a body's jurisdiction?}}
\end{description}

\subsection{Research Approach}

A UK-focused case study-based approach has been adopted for this project. As is common in case study-based research, many alternative case studies could have been chosen; furthermore, the cases are illustrative rather than comprehensive in terms of the available case studies or challenges. The cases were evaluated to articulate the progress made and highlight opportunities for future developments in education and practice. 

The first set of cases that are considered are the current situation in terms of recommendations from a sample of relevant PSRBs, together with the published evidence regarding compliance with these recommendations; this is contextualised by the wider policy context. Together these provide a context to the ongoing enhancement initiatives in the area of cybersecurity education.

In order to address {\emph{RQ1}} and {\emph{RQ2}} a number of specific case studies pertaining to challenges of delivering cybersecurity to undergraduate computer science students are evaluated. Namely: an industry problem with evident cybersecurity implications; the current state of educational resources with respect to cybersecurity; and the challenges evident in the UK related to the recruitment and retention of suitably qualified academic staff in the cyber security area. 

Finally, in order to address {\emph{RQ3}}, case studies are evaluated related to the challenges and successes of PSRB accreditation of cybersecurity in undergraduate computer science in one jurisdiction (the UK). As discussed, mandating cybersecurity within PSRB accreditation in this jurisdiction is at a reasonably mature stage.

\subsection{Current State of Play}

The ACM/IEEE-CS Joint Task Force on Computing Curricula~\cite[p.~97]{ACM2013a} takes a distinct view on the Knowledge Areas (KAs) related to {\emph{RQ2}}:

\begin{quote}
``{\emph{The Information Assurance and Security KA is unique among the set of KAs presented here
given the manner in which the topics are pervasive throughout other KAs.}}''
\end{quote}

It proposes nine ``core'' hours and 63.5 distributed across the other KAs. Nevertheless, the situation on the ground in the USA is different~\cite{Ackerman2019a}:

\begin{quote}
``{\emph{Universities suffer shortcomings, as well. Roughly 85 of them offer undergraduate and/or graduate degrees in cybersecurity. There is a big catch, however. Far more diversified computer science programs, which attract substantially more students, don't mandate even one cybersecurity course.}}''
\end{quote}

The UK situation is distinctly different: 61\% of UK courses offer mandatory cybersecurity content, and this research was based on web scraping~\cite[Table 1]{Ruiz2019a}. As such it represents a lower bound since not all coverage will necessarily be clearly articulated in publicly available documentation online. It is at least plausible to attribute this difference to differences in the accreditation regimes, as the external circumstances, governmental pressures, and professional body/learned society curricula are all similar.

\begin{description}
\item[UK] BCS, The Chartered Institute for IT (BCS) has had a requirement to include information security in the curriculum since 2010, and has expected coverage of an agreed cybersecurity syllabus since 2015 (Table \ref{table:1}), with the result that all accredited universities should be compliant by 2020 (due to the five-year cycle). More precisely, accredited degrees have been expected to demonstrate coverage of ``{\emph{2.1.9 Knowledge and understanding of information security issues in relation to the design, development and the use of information systems}}'' \cite[p.~30]{BCS2018a} since 2010 with an enhanced cybersecurity related definition of what this entails since 2015 \cite[p.~17--18]{BCS2018a}.
\item[USA] The Association of Computing Machinery (ACM) has equally had cybersecurity (IAS: ``Information Assurance and Security'') in the curriculum since 2013~\cite{ACM2013a}, but it is not the accrediting body. The Accreditation Board for Engineering and Technology (ABET) is, and is requiring IAS with effect from the 2019-20 cycle (self-study reports due 1 July 2019): more precisely \cite[Table 3]{Oudshoornetal2018a} ``{\emph{The computing topics must include: \dots{} Principles and practices for secure computing\dots}}''. This should mean that  all accredited universities should be compliant by 2025 (due to their six-year cycle).
\end{description}

\section{Challenges: ({\emph{RQ1}} and {\emph{RQ2}})}

The ACM Computer Science Curricula Recommendations~\cite{ACM2013a} states three Tier-1 and six Tier-2 hours for ``Information Assurance and Security'', but this is the ``IAS-only'' topics, and ACM expects 32 Tier-1 and 31.5 Tier-2 hours for IAS topics embedded in other Knowledge Areas.

The UK's official body of knowledge resource, the CyBOK project~\cite{Bristol2019a}, has produced reference documentation for some (as of June 2019: five final, seven for comment, out of a planned total of 19) knowledge areas, which are useful references for the experienced educator looking for a definition or characterisation, but a long way from being a textbook (which is not their aim). 

\subsection{Industry Standards: PCI DSS}\label{sec:PCIDSS}

Is teaching cybersecurity different from other computing topics? Lecturing is probably not the best way; how should we use real-life case studies (e.g. British Airways \cite{Barth2018a}) and authentic assessment? Recent work suggests that there are benefits from teaching this discipline area in a more practical manner \cite{Weiss:2013:THC:2527148.2527180}; one would be by the inclusion of pertinent cybersecurity standards in the curricula.

The {\emph{Payments Card Industry Data Security Standards}} (PCI DSS)~\cite{PCI2018b} is one such standard, underpinning all processing of credit/debit cards globally. Nevertheless, they are very rarely mentioned in generalist computer scientist courses. This would not matter so much if everyone handling payments data were sent by their employers on an effective PCI DSS course. However, the payments business is now so spread across websites, often run by small and medium enterprises (SME), or non-specialists.  Even larger enterprises are not immune: the 2018 British Airways breach was caused by a failure to adhere to PCI DSS in the website maintenance~\cite{Barth2018a}. 

It is an interesting question as to whether standards such as PCI DSS should be addressed within degree courses (clearly degree courses can never cover all standards) or whether they should be addressed in professional training courses. However it appears the current situation is not ideal from the perspective of industry (or users of systems) and the inclusion of key standards could be seen as a valuable enhancement activity to the coverage of cybersecurity within generalist computer science courses.

\subsection{Educational Resources}\label{sec:EDResource}

\subsubsection{SQL Injection}\label{sec:SQL}

It is 15 years since \cite{Guimaraesetal2004} wrote ``{\emph{All the topics listed above should be presented in the first Database Course}}'', and the first such topic was SQL injection \cite{SPIDynamics2002,Anonymous2018b}. SQL injection as an attack has been around for twenty years \cite{HornerHyslip2017a}, has its own cartoon and website\footnote{See: \url{https://xkcd.com/327/} (dating back to 2007, according to the Internet Archive) and \url{http://bobby-tables.com/}}. Nevertheless SQL injection is still a major weakness: number one in the Open Web Application Security Project (OWASP) Top 10 \cite{OWASP2017a}, and has been in the Top 10 since at least 2003. CyBOK states ``{\emph{\ldots a wide range of attack techniques for exploiting SQL injection or script injection are known and documented.}}''~\cite{Bristol2018a}.

Clearly such a major weakness should be well-taught; in general it is hard to determine what is actually delivered as part of a specific degree programme, but a reasonable proxy for this is the content of recommended textbooks. This was the rationale for a 2019 analysis of database textbooks used by 44 of the top 50 computer science departments in the USA~\cite{Drop2019}.  There were seven such books, but three books accounted for the 36 of the 44 universities. Five of the seven (30 of the 44) had no mention of SQL injection. On the other two, the more popular one has a seriously flawed discussion\footnote{``{\emph{However, the paper implies that using parameters is equivalent to using a function to add escape characters around user input. This is incorrect, as using parameters allows SQL statements to be pre-compiled, and prevents any user input from being interpreted as code, while escaping user input is not recommended as a sole defense since imperfect escape functions can easily be subverted.}}''~\cite{Drop2019}}, and the other, while generally excellent, had a presentational problem\footnote{``{\emph{However, the fact that the first example should not be used is not discussed until two pages after the example in the text, and is not mentioned at all in the caption or on the page where the figure appears. This means a student who is skimming the text looking for an example to modify for their own code could simply copy the code that first appears in the example, without being aware that this is in fact an example of what they shouldn't do.}}''~\cite{Drop2019}}. This oversight is not limited to textbooks: although Wikipedia has an article on SQL Injection, it was not linked from the SQL page itself at the time of writing\footnote{Fixing this is a suggestion for future work.}.

\subsubsection{The Case of Java}\label{sec:Java}

A recommendation for future work is a comprehensive survey equivalent to \cite{Drop2019} for  Java textbooks. Indeed, many such books go nowhere near security applications.  But this means that the programmer who has to implement security is left to the documentation of the package/API being used, and to informal resources. \cite{Mengetal2018a} analysed 503 cybersecurity-related postings on the popular Stack Overflow online resource. 53\% were about the Spring Security framework\footnote{\url{https://projects.spring.io/spring-security/}}, dominated by authentication (45\%). The discussion \cite[\S4.3.1]{Mengetal2018a} of cross-site request forgery (CSRF) is especially worrying.  By default, Spring implicitly enables protection against this. But all the accepted answers to CSRF-related failures simply suggested disabling the check. There were no negative comments about this, and indeed a typical response is ``{\emph{Adding}} \verb!csrf().disable()!
{\emph{{solved the issue!!! I have no idea why it was enabled by default}}''. As of writing, there were no negative comments about this disabling of a vital security feature. This research was further developed by \cite{Chenetal2019a}  (and popularised in a security community in \cite{Zorz2019a}). Their first finding was:

\begin{quote}
``{\emph{644 out of the 1,429 inspected answer posts
(45\%) are insecure, meaning that insecure suggestions
popularly exist on SO. Insecure answers dominate, in
particular, the SSL/TLS category}}'' [355 insecure versus 150 secure, i.e. $>70$\%].
\end{quote} 

\subsubsection{Android}\label{sec:Android}

Another recommendation for future work is a comprehensive survey equivalent to \cite{Drop2019} for Android textbooks; \cite{Fischeretal2017a} looked specifically at the use of resources from Stack Overflow in Android applications. The key finding was:

\begin{quote}
``{\emph{We found that 15.4\% of all 1.3 million Android applications
contained security-related code snippets from
Stack Overflow. Out of these 97.9\% contain at least one
insecure code snippet.}}''
\end{quote}

Two caveats (in opposite directions) should be noted. The labelling was conservative, in that snippets were only labelled as insecure if that was demonstrable, and, for example, mere use of outdated SSL/TLS was not automatically deemed insecure. On the other hand, the insecure snippet might have been used in a way that did not expose the insecurity. The uncritical reading of Stack Overflow was also noted in \cite[Slide 29]{Votipkaetal2019a}. Their key recommendation~\cite[Slide 32]{Votipkaetal2019a} was ``{\emph{Improve documentation: Clarify what you can(not) copy/paste}}''. 

\subsubsection{Agile}

A further recommendation for future work is a comprehensive survey equivalent to \cite{Drop2019} for ``Agile'' textbooks. Many authors have found disconnects between Agile practices and secure software development: notably \cite{Bartsch2011a} for small projects and \cite{vanderHeijden:2018:EPS:3239235.3267426} for large projects. Agile's preference for functionality over non-functional requirements is clearly displayed in practice. \cite{Naiakshinaetal2017a} asked 20 student developers to imagine they were part of a team working on creating a social networking site for our university and to implement a password storage mechanism for this. 10 (``primed'') were explicitly told that the storage had to be secure and 10 (``unprimed'') were not. None of the unprimed ones implemented any security. This would also reaffirm the point raised in section~\ref{sec:PCIDSS}, regarding the effective use of industrial case studies and authentic assessment approaches.

\subsubsection{Informal Resources}\label{sec:informal}

The web abounds with informal resources, such as tutorials and code snippets. How good are these, and how good are people at using these? This has been looked at by \cite{Unruhetal2017a}, taking the top five search results from Google for six queries. Of these 30 tutorials, six had SQL injection weaknesses, and three had Cross-Site Scripting\footnote{Number 7 in OWASP's Top Ten \cite{OWASP2017a}.} weaknesses. Searching for these fragments in PHP projects on GitHub found 820 instances of these fragments, of which 117 were verified manually to be vulnerable --- 80\% of which were vulnerable to SQL injection. Some students clearly make use of these resources; thus a recommendation of future work is to explore and evaluate students' (and indeed others') use of such informal resources. 

\subsection{Faculty}

It is well known that cybersecurity skills are in short supply, in both industry~\cite{Ackerman2019a} and academia~\cite{schneider2013}. For example:

\begin{quote}
{\emph{Research into the state of IT conducted annually by Enterprise Strategy Group (ESG)~\cite{ESG:2018} has revealed that the skills gap in information security continues to widen and has doubled in the past five years. In 2014, 23\% of respondents to the survey stated that their organisation had a problematic shortage of information security skills. This had climbed to 51\% at the beginning of this year. Clearly, this is an issue which is being felt across many industries and organisations, and is a concern which extends beyond IT leadership into the boardroom.}}~\cite{Page2018a}
\end{quote}

The ESG survey is international, but ESG have confirmed that the UK figures are very similar. In the UK, there has been a resurgence of job adverts to recruit academic staff with specialisms in cybersecurity over the past three years. The demand for cybersecurity skills in industry makes it difficult for academia to attract academics with knowledge, practical experience, research background and academic aspirations. As universities expand their cybersecurity provision it is not uncommon to find multiple jobs advertised at the same time. Recent example have included a professor of cybersecurity, two senior academic positions and two junior academic positions in one advert. There are other examples in the UK of cybersecurity lecturing jobs remaining unfilled for longer than a year; there are also examples of cybersecurity research groups moving en masse from one university to another.

\section{Accreditation by PSRBs ({\emph{RQ3}})}

In the UK --- as in most jurisdictions --- higher education provision addresses general computer science and specialist cybersecurity courses. A significant number of undergraduate and postgraduate programmes are available in both the areas of computer science and cybersecurity (and closely related fields, such as computer security, digital/computer forensics, etc). The UK's Universities and Colleges Admissions Service (UCAS) lists over 40 higher education institutions providing undergraduate qualifications related to cybersecurity for entry in September 2019. An even larger number of UK institutions provide study opportunities related to more general undergraduate computer science: UCAS lists 246 providers for undergraduate programmes for the 2019/2020 academic year. 

Accreditation has evolved to directly addresses the cybersecurity challenges in both general computer science programmes and specialist cybersecurity programmes. In the UK, accreditation in the broad computing area is being performed by a few organisations with overlapping interests and priorities. These include:

\subsection{National Cyber Security Centre (NCSC)}

The NCSC is a UK Government organisation tasked with enhancing the cybersecurity of the UK. The NCSC publishes and accredits to a number of cybersecurity standards~\cite{NCSC2017}; these standards are linked to the ACM recommendations for curricula~\cite{ACM2013a}. To date, the major focus of NCSC accreditation has been upon Masters degrees specialising in cybersecurity. More recently the NCSC has also began accrediting integrative masters programmes, undergraduate degrees in cybersecurity, and computer science degrees with a significant cybersecurity focus.

The NCSC have accredited 15 cybersecurity MSc programmes with a further 11 provisional accredited, 3 integrated masters cybersecurity programmes and 1 cybersecurity degree programme, with a further 2 provisionally accredited. Hence the extent of accreditation is currently reasonably limited in terms of reach to computer science programmes. This appears to be a positive initiative that will further develop over the next few years, especially with increasing industry engagement in this area.

\subsection{Not-for-Profit Organisations}

Tech Partnership Degrees is a not-for-profit organisation that provides endorsements to higher education programmes with specific curricula elements aimed at job market requirements. One of the curricula elements is related to cybersecurity. Tech Partnership Degrees have a specialist scope, endorsing programmes in the named areas of ``IT Management for Business'' and ``Software Engineering for Business''. Tech Partnership Degrees currently endorse 14 IT Management for Business Programmes and five Software Engineering for Business Programmes. As such, Tech Partnership Degrees currently have limited impact upon more general computer science education and none upon specialist cybersecurity education. 

The Institute of Coding\footnote{\url{https://instituteofcoding.org}} is a not-for-profit entity, led by the University of Bath, that intends to enhance how digital skills are developed in higher education in the UK~\cite{Davenportetal2019a}. This is likely to include cybersecurity related skills. Like the Tech Partnership, the focus is upon job market requirements. Additionally the Institute is looking at potentially endorsing the demonstrable capabilities of graduates as shown by their university studies, work experience and work placements. Given the size of this initiative this potentially has a significant role to play however at time of writing it is clearly a work in progress.

\subsection{Professional Bodies and Learned Societies}

Both BCS, The Chartered Institute for IT (BCS) and the Institution of Engineering and Technology (IET) accredit programmes in the general area of computer science and the more specialist area of cybersecurity discipline areas. The accreditation provided by these institutes are underpinned by international initiatives such as the Washington Accord\footnote{\url{http://www.ieagreements.org/accords/washington/}.} and the Seoul Accord\footnote{\url{https://www.seoulaccord.org/}.}. These memoranda support the internationalising of the curriculum and promote consistency and parity in computer science education globally.  These professional bodies are also UK charities and hence have responsibilities for public good which extends beyond short term job market needs~\cite{Stensaker2006,Mutereko2017}. Both the BCS and the IET have a long history of expecting coverage of environmental factors within the programmes they accredit. The BCS has for a number of years been expecting significant coverage of what it terms  {\emph{Legal, Social, Ethical and Professional Issues}}~\cite{Brooke2018}. Clearly cybersecurity has been and continues to be part of these expectations.

\section{Case Study: BCS Accreditation}

In recent years the BCS has evolved its accreditation practices to promote and mandate the inclusion of cybersecurity within the programmes the body accredits. The timeline which this process has followed in presented in Table~\ref{table:1}, and will be discussed in more detail in the following case study.

  \begin{table}[h!]
  \begin{tabularx}{\textwidth}{ |X|X| }
    \hline
    IV. ACCREDITATION ({\emph{RQ3}}) - A. Developing Expectations &  \\ \hline
    UK Government Cybersecurity Strategy \cite{ukcyberstrategy:2016} & November 2011 \\ \hline
    Three workshops of a consortium of industry, academia and government bodies -- led by CPHC and  (ISC)$^2$ -- leading to the development of cybersecurity learning guidelines to be embedded into BCS accredited UK computer science and IT-related degrees~\cite{CPHCISC2}  & 2013 to June 2015 \\ \hline
    UK Government report Cybersecurity Skills, Business Perspectives and Government's Next Steps Report Released \cite{UKCabinetOffice2014} & March 2014  \\ \hline
    Council of Professors and Heads of Computing (CPHC) Identifies Cybersecurity as one the top 3 concerns in Computing & April 2014 \\ \hline
    Joint Development of White Paper from CPHC and The International Information Systems Security Certification Consortium (ISC)$^2$ \cite{CPHCISC2014} & April -November 2014 \\ \hline
    Extended Cybersecurity Criteria included in BCS Accreditation Guidelines \cite{BCS2018a}& June 2015 \\
    \hline
    IV. ACCREDITATION ({\emph{RQ3}}) - B. What does the BCS tell Universities? & \\ \hline
    Cybersecurity Principles Roadshow & March-April 2016 \\ \hline
    IV. ACCREDITATION ({\emph{RQ3}}) - C. Accreditation - what progress has been made? &  \\ \hline
    All visited institutions expected to be fully compliant 
 (BCS follow a five-year accreditation cycle)  & September 2020\\ \hline
  \end{tabularx}
\caption{Timeline of the development of cybersecurity expectations in the UK}
  \label{table:1}
  \end{table}

\subsection{Developing Expectations}

Internationally the expectations regarding both the breadth and depth of the expected cybersecurity coverage has been the subject of much discussion, debate and analysis. Like many governments, the UK Government has actively been seeking ways to address this~\cite{UKCabinetOffice2014,ukcyberstrategy:2016}. In parallel to the work completed by the ACM~\cite{ACM2013a} in the USA, considerable effort in the UK have been taken to ensure industry, higher education, government and the relevant professional bodies collaborated on a set of guidelines which are to the benefit of the various stakeholders and wider society~\cite{Irons2016}. In 2013, an initiative was set up by (ISC)$^2$, CPHC (the representative body of UK computer science departments) and the UK Cabinet Office to examine embedding cybersecurity into undergraduate degrees in the UK. Three workshops in 2013 and 2015 attempted to define the principles of cybersecurity education and proposed a framework for embedding these principles in UK computer science curricula. Attendees at the workshops included industry, professional bodies, UK government departments and more than 30 universities that offer undergraduate computer science degrees. This work initially lead to a white paper related to a proposal in the form of a white paper in 2014~\cite{CPHCISC2014}, followed by a set of guidelines in 2015~\cite{CPHCISC2}. The BCS agreed to adopt the outputs into their accreditation criteria. This was the first time that cybersecurity has been extensively referenced within accreditation criteria for computing and IT-related degrees. The fact that cybersecurity is included as a component of the BCS accreditation criteria reflects the importance placed on cybersecurity and the expectation that all computing graduates should have knowledge and skills in cybersecurity as they move towards chartered status.

The produced reference guidelines (``{\emph{Cybersecurity Principles and Learning Outcomes}}'') \cite{CPHCISC2} established a baseline of common knowledge, example learning outcome domains for cybersecurity within the computer science courses and guidance on embedding the concepts. The document provides specific guidance for embedding and enhancing relevant cybersecurity principles, concepts and learning outcomes within their undergraduate curricula. The document suggested five areas of coverage 

\begin{itemize}
    \item Information and risk;
    \item Threats and attacks;
    \item Cybersecurity architecture and operations;
    \item Secure systems and products; and 
    \item Cybersecurity management.
\end{itemize}

The ambition of this approach was to influence the curricula of all programmes seeking accreditation (regardless of the precise discipline area); it was not intended to be prescriptive or stifle innovation, however it is intended to promote curricula that would benefit the students upon programmes, their future employers and wider society.  In this context this is realised as an expectation cybersecurity is an inclusion in all degrees accredited by the BCS. e.g. the expectation for coverage is true for computer science as well as cybersecurity programmes. Two criteria are expected to be covered by all programmes seeking accreditation. These are:

\begin{quote}
``{\emph{2.1.6 Recognise the legal, social, ethical and professional issues involved in the exploitation of computer technology and be guided by the adoption of appropriate professional, ethical and legal practices}}'''~\cite{BCS2018a}
\end{quote}

\begin{quote}
``{\emph{2.1.9 Knowledge and understanding of information security issues in relation to the design, development and the use of information system}}''~\cite{BCS2018a}
\end{quote}

Additionally, programmes seeking Chartered Information Technology Professional (CITP) accreditation also have to cover:

\begin{quote}
``{\emph{3.1.2 Knowledge and understanding of methods, techniques and tools for information modelling, management and security}}''~\cite{BCS2018a}
\end{quote}

In the context of BCS accreditation, these requirements imply an exit standard that all students on a programme must be able to demonstrate irrespective of the option choices they have made. This means an institution applying for accreditation is expected to provide evidence that the criteria are taught and assessed in a non-trivial manner, to and by all students upon the programme seeking accreditation. An institution is expected to provide evidence in the form of programme and module specification documentation and example assessment specifications (coursework and examinations). These criteria and the expectation that they are taught and assessed has been present for a number of years.

The relevant BCS accreditation (criteria 2.1.6, 2.1.9 and 3.1.2, as presented above) is not prescriptive, but encourages institutions to embed cybersecurity teaching across a range of subject areas in the computer science curriculum such as programming, software design, databases, networking, architecture. In addition there an expectation that there is significant coverage of cybersecurity principles and fundamentals -- either as a stand alone module or as a significant component(s) of other modules. This approach differs from the ACM approach~\cite{ACM2013a}, where the expectations are more explicit and the curriculum expectations are specified at a more granular level.

\subsection{What Are Universities Told?}

The agreed {\emph{Cybersecurity Principles and Learning Outcomes}}~\cite{CPHCISC2} were discussed with the wider education community, led by CPHC. A series of workshops took place in 2015 which presented the rationale for embedding cybersecurity in the curriculum of computer science degrees. The workshops included case studies from universities who had embedded cybersecurity into their computer science curricula illustrating different approaches to implementation. The workshops had 102 attendees from the academic computer science community representing 60 UK Universities. 

The {\emph{BCS Guidelines on Course Accreditation}} are published online~\cite{BCS2018a}, with the BCS publishing any changes that have been made~\cite{BCS2018b}. When changes are made, the BCS communicates the changes by email and in writing to all the BCS Educational Affiliates, that is all the institutions that seek accreditation from the BCS. The expectations for cybersecurity were extended in the June 2015 version of the guidelines for consideration at accreditation visits that took place from September 2015 or later.

This change to the accreditation guidelines is now in an implementation period. The accreditation process adopted by the BCS is cyclic in nature; formally, the cycle is five years in duration. The new expectations have been implemented as follows: to ensure continuous accreditation, accreditation visits are normally scheduled every five years; at the time of the next visit in this accreditation cycle, accreditation is conditional upon an institution having considered the guidelines and either adjusted the curriculum to meet the new expectations or have a formal plan in place for when and how adjustments will be made.  It is anticipated that from 2020 the expectation will be all accredited programmes have the new expectations fully embedded.

In the year prior to an accreditation visit, institutions are invited to a briefing from the BCS. The intention of the briefing is to help ensure accreditation visits are successful from the perspective of both the BCS and the institution. The briefings take place virtually and includes a summary of the process, discussion of recent changes, guidance regarding the application and a summary of common issues that are being seen in other institutions. Significant opportunity for seeking clarification is provided. One of the issues often highlighted is that not all institutions have yet evolved their programmes to fully address the increased expectation for cybersecurity. This is resulting in accreditation being contingent upon an institution taking action to address this short fall or in some cases the withdrawal of accreditation. A number of institutions are in the process of adjusting their curricula to meet the new expectations. In this case, the BCS notes the changes to programme design, the outputs from which will be scrutinised at the next accreditation visit.

\subsection{Accreditation: What Progress Has Been Made?}

This initiative is a collective attempt to formally include cybersecurity in all BCS Accredited programmes. Some of these programmes will be specialist cybersecurity programmes, however the majority will take a different emphasis; this is a work in progress. A full cycle of accreditation visits has not yet taken place following the adjustment to the BCS Guidelines~\cite{BCS2018b}. What is being observed is the majority of visited institutions have now either adjusted their curricula to extend the coverage of cybersecurity or have a plan in place to do so. However, a minority are requiring encouragement to do so.

From the start of the Autumn 2015 term, up to and including the Autumn 2018 term, the BCS have carried out 70 accreditation visits (including four international visits). The BCS identified action was required to address concerns related to cybersecurity at 16 institutions; thus, 54 institutions were already delivering cybersecurity in keeping with the BCS expectations.

Long-term actions were expected from 12 institutions (six in 2015/16, three in 2016/17 and three in Autumn 2018) who were awarded `{\emph{At Threshold}}' judgments. Ten of these judgments were across all programmes; one was specifically against a generalist masters programme only. This indicates that the BCS will expect adjustments to have taken place before the next accreditation visit. As indicated earlier, this was commonly the case that adjustments had been made to approved programmes of study, however the adjusted modules had not yet been delivered so the evidence base was incomplete in terms of how cybersecurity was assessed.
 
Short terms 90 Day Responses where required from four institutions; the outcomes of these actions were as follows: ({\emph{i}}) of the 11 UG programmes involved all were approved `{\emph{At Threshold}}'; ({\emph{ii}}) of the nine UG programmes involved, eight were approved and one refused; ({\emph{iii}}) of the five UG programmes involved, all approved `{\emph{At Threshold}}'; and ({\emph{iv}}) of the 3 UG programmes involved, all 3 were refused.

 
Good practice was identified at one university by the commendation:

 \begin{quote}
``{\emph{The second year project provides an opportunity for exploring security aspects in depth with an industrial use case.}}''
\end{quote}

In summary, this shows that many UK institutions have now embedded cybersecurity in their provision, a number are in the process of doing so and a minority have chosen not to. Clearly not all institutions in the UK necessarily have to apply for PSRB accreditation, or apply for accreditation for all their programmes, but even so this is significant evidence of inclusion of cybersecurity to an agreed standard.

\section{Conclusions and Future Work}

The work presented in this paper is a first step towards better understanding the nature, design, structure and assessment of cybersecurity education in the UK, through the lens of accreditation. It is clear that PSRB accreditation is having a positive effect on universities, supporting wider national policy imperatives. It is also clear that there is a significant need for mobilising the international computer science academic community -- alongside some of the existing initiatives presented at the start of this paper -- to continue this focus on cybersecurity education, and provide international comparators, portability and sharing of best practice. With regards to the three research questions:

\begin{description}
\item[RQ1:] {\emph{What cybersecurity is taught and what cybersecurity should be taught to the general computer science students?}}

The guidelines from both ACM and BCS are good for general education. However, the most important item would seem to be an attitude of caution with respect to both offline (\S\ref{sec:SQL}) and online (\S\ref{sec:Java}) resources. 

\item[RQ2:] {\emph{Should cybersecurity be taught stand-alone or in an integrated manner to general computer science students?}}

The recommendation in \cite[p. 98]{ACM2013a} that cybersecurity be taught largely through other Knowledge Areas is, in abstract, a good idea.  However, in the current state of education resources (sections \S\ref{sec:EDResource}) may be a counsel caution in this approach.  It is more important that issues like SQL injection \cite{Drop2019} or correct use of SSL/TLS \cite{Chenetal2019a} be taught somewhere than that they not be taught at all. Nevertheless, it is wrong for a complete curriculum to ignore cybersecurity issues for example:

\begin{itemize}
	\item teach SQL without teaching SQL Injection~\cite{Drop2019};
	\item teach ``web programming'' without teaching Cross-Site Scripting \cite[(XSS)]{OWASP2017a};
	\item ignore the  impact of GDPR upon systems development;
	\item uncritically teach the use of Agile development to develop secure systems.
\end{itemize}

The BCS-identified good practice of exploring cybersecurity via a project referencing Industry Standards, possibly PCI DSS, is commended (\S\ref{sec:PCIDSS}).

\item[RQ3:] {\emph{Can accreditation by PSRBs enhance the provision of cybersecurity within a body's jurisdiction?}}

Accreditation (as practised by BCS in the UK) is a valuable tool in improving the standard of cybersecurity teaching, and disseminating good practice, and should continue this. The time-lag in adoption of cybersecurity in the USA seems to be caused by the accreditation differences more than any other factors.
\end{description}

We also have the following specific recommendations:

\begin{enumerate}
\item Database courses should look carefully at the security aspects of the texts they use, and the examples they quote, on the lines of \cite{Drop2019}.
\item Web Programming courses should do the same, with emphasis on the avoidance of Cross-Site Scripting, and, for production use, the use of a suitable framework that has Cross-Site Request Forgery protection\footnote{Most modern frameworks do, but this is not often discussed when looking at the advantages of frameworks. Even Mozilla's tutorial (\url{https://developer.mozilla.org/en-US/docs/Learn/Server-side/First_steps/Web_frameworks}) is silent.}.
\item A study similar to \cite{Drop2019} is recommended for future work focusing on Java development, Android development and Agile development, and potentially other areas of the curriculum, to better understand how modern textbooks support key areas of the curriculum.
\item All computer science courses should emphasise that informal resources should come with a ``security health warning'': see sections \ref{sec:informal} and \ref{sec:Java}. One should probably use the data from \cite{Chenetal2019a}: ``{\emph{If you pick up a SSL/TLS answer from Stack Overflow, there's a 70\% chance it's insecure}}''.
\item Finally, further work exploring and evaluating effective pedagogical and assessment approaches for cybersecurity education, in a similar vein to what has been seen with computer science education more generally (e.g.~\cite{davenport-et-al:latice2016}).
\end{enumerate}

\section*{Acknowledgements}

The authors wish to thank Sally Pearce, Academic Accreditation Manager at BCS, The Chartered Institute for IT for supplying the summary information related to accreditation of UK degree programmes. Many people, accreditors and accredited, have contributed to improving the standard of cybersecurity teaching in the UK, and spreading good practice.  All authors' institutions are members of the Institute of Coding, an initiative funded by the Office for Students (England) and the Higher Education Funding Council for Wales.

{\emph{N.B.}} The first two authors are current Vice-Presidents of the BCS, and the third and fourth are past and present Chairs of the BCS Academic Accreditation Committee, providing significant personal insight into the UK's national accreditation policy and procedures.

\bibliographystyle{unsrt}
\bibliography{FIE2019}

\end{document}